\documentclass[aps,pra,twocolumn,amsmath,amssymb,nofootinbib,showpacs]{revtex4}
\usepackage[english]{babel}
\usepackage{times}

\usepackage{latexsym}
\usepackage{graphics}
\usepackage{graphicx}
\usepackage{subfigure}
\usepackage{epsfig}

\newcommand{\infig}[2]{\begin{center}
                                    \mbox{ \epsfxsize #1 \epsfbox{#2.eps}}
                                      \vspace{-0.8cm}
                                    \end{center}}

\renewcommand{\acknowledgements}{\section*{ACKNOWLEDGEMENTS}}

\newcommand{\eprint}[1]{#1}

\newcommand{\ie}{{i.e. }}

\newcommand{\vect}[1]{{\bf #1}}

\newcommand{\erfc}{\textrm{erfc}}
\newcommand{\sinc}{\textrm{sinc}}

\newcommand{\Vopt}{V}
\newcommand{\tVopt}{\widetilde{V}}
\newcommand{\Vs}{V_\textrm{\tiny R}}

\newcommand{\corr}{C}
\newcommand{\tcorr}{\widetilde{C}}
\newcommand{\gecr}{G}

\newcommand{\muTF}{\mu_\textrm{\tiny TF}}
\newcommand{\LTF}{L_\textrm{\tiny TF}}

\newcommand{\sigmas}{\sigma_\textrm{\tiny R}}
\newcommand{\tsigmas}{\widetilde{\sigma}_\textrm{\tiny R}}
\newcommand{\sigmaG}{\sigmas}
\newcommand{\sigmaGline}{\overline{\sigma}_\textrm{\tiny R}}

\newcommand{\asc}{a_\textrm{sc}}
\newcommand{\goned}{g_\textrm{\tiny 1D}}
\newcommand{\gdD}[1]{g_{\tiny #1\textrm{D}}}

\begin{document}

\title{Smoothing effect and delocalization of interacting Bose-Einstein condensates in random potentials}

\author{L. Sanchez-Palencia}
\affiliation{
Laboratoire Charles Fabry de l'Institut d'Optique,
CNRS and Univ. Paris-Sud,
Campus Polytechnique, 
RD 128, 
F-91127 Palaiseau cedex, France}
\homepage{http://www.atomoptic.fr}

\date{\today}

\begin{abstract}
We theoretically investigate the physics of 
interacting Bose-Einstein
condensates at equilibrium in a weak (possibly random) potential. 
We develop a perturbation approach to derive the condensate
wave function for an amplitude of the potential smaller than the chemical potential of
the condensate and for an arbitrary spatial variation scale of the potential. 
Applying this theory to disordered potentials, we find in particular that, if the healing 
length is smaller than the correlation length of the disorder, the condensate assumes 
a delocalized Thomas-Fermi profile. 
In the opposite situation where the correlation length is smaller
than the healing length, we show that the random potential can be
significantly smoothed and, in the mean-field regime, the condensate 
wave function can remain delocalized, even for very small correlation 
lengths of the disorder.
\end{abstract}

\pacs{03.75.Hh;79.60.Ht}

\maketitle

%%%%%%%%%%%%%%%%%%%%%%%%%%%%%%%%%%%%%%%%%%%%%%%%%%%%%%%%%%%%%%%%%%%%%%
\section{Introduction}
\label{sec:intro}

Ultracold atomic gases are currently attracting a lot of attention
from both experimental and theoretical viewpoints.
Taking advantage of the recent progress in cooling and trapping of neutral 
atoms \cite{nobel1997}, dilute atomic Bose-Einstein condensates
(BECs) \cite{nobel2001} and degenerate Fermi gases (DFGs) 
\cite{truscott2001,schreck2001,hadzibabic2002,roati2002} 
are now routinely produced at the laboratory. 
Using various techniques, space-dependent potentials can be designed 
almost on demand in these systems. For example, one can produce 
periodic \cite{grynberg2000,verkerk1992},
quasi-periodic \cite{guidoni1997,quasi2005,schulte2005},
or random potentials \cite{horak1998,grynberg2000bis,lye2005,clement2005,fort2005,clement2006} 
by using optical means.
For these reasons and due to unique control and analysis possibilities,
ultracold gases constitute a favorite playground for revisiting standard problems 
of condensed matter physics (CM) \cite{anglin2002,jaksch2005,giamarchi2006,lewenstein2006}.

Most current experiments with BECs lie in the mean-field regime where the Bose gas
is described by a single
wave function, $\psi$, governed by the 
(nonlinear) Gross-Pitaevskii equation \cite{pitaevskii2004}.
Due to the interplay between the kinetic energy term and the interaction term,
it is usually difficult to derive the exact solution of this equation.
The importance of interactions can be characterized by 
the {\it healing length}, $\xi$, which defines the typical distance below which the spatial 
variations of $\psi$ significantly contribute to the energy of the BEC, {\it via} 
the kinetic energy term \cite{pitaevskii2004}.
In the Thomas-Fermi regime (TF), \ie when $\xi$ is significantly smaller 
than the typical variation scale, $\sigmas$, 
of the potential, $V(\vect{r})$, to which the BEC is subjected, the kinetic term is 
negligible and the BEC density
simply follows the spatial variations of the 
potential\footnote{This is standard in the case of a harmonic confinement,
$V(\vect{r})=m\omega^2\vect{r}^2/2$. Although, there is no intrinsic
typical variation scale,
one can define $\sigmas$ as $m\omega^2\sigmas^2/2 = \mu$, \ie
$\sigmas=\LTF$, the usual TF half size of the condensate and the validity
of the TF regime reads $\xi \ll \LTF$.
For periodic, quasi-periodic or random potentials, $\sigmas$ is the spatial period 
or the correlation length (see section~\ref{sec:random} for details).}:
%+++++++++++++++++++++++++++++++++++++++++%
\begin{equation}
|\psi (\vect{r})|^2 \propto \mu - V(\vect{r}) ~.
\label{eq:TF0}
\end{equation}
%+++++++++++++++++++++++++++++++++++++++++%
In the opposite situation ($\xi > \sigmas$), the kinetic term should be taken into account 
and the exact BEC wave function usually cannot be found analytically.

Besides a general interest, the question of determining the BEC wave function for an 
arbitrary ratio $\sigmas/\xi$ has direct applications to the case where $V(\vect{r})$ 
is a random potential. The physics of quantum systems in the presence of disorder
is central in CM \cite{nagaoka1982,ando1988,vantiggelen1999},
owing to unavoidable defects in `real-life systems'. 
One of the major paradigms of disordered quantum systems is due to Anderson 
who has shown that the eigenstates of single 
quantum particles in arbitrary weak random potentials can be localized, 
\ie $\psi$ shows
an exponential decay at large distances\footnote{In 1D and 2D systems, all 
eigenstates are usually localized while in 3D, they are localized below the
so-called {\it mobility edge}.} \cite{anderson1958}. 
Recent experiments have studied the onset of strong or weak localization effects of 
light waves \cite{Wiersma_Nature1997,labeyrie1999} and 
microwaves \cite{Dalichaouch_Nature1991,Chabanov_Nature2000}. 
Ultracold matter waves are also widely considered 
as promising candidates to investigate Anderson localization in random 
\cite{damski2003,gavish2005,paul2005} 
or quasi-random structures \cite{quasi2005,damski2003,roth2003} and more
generally to investigate the effects of disorder in various quantum systems
(for a recent review, see Ref.~\cite{ahufinger2005} and references therein).
It is expected that the dramatic versatility of ultracold gases would
allow us for a direct comparison with theoretical studies of quantum disordered systems.

A key peculiarity of BECs is that interactions usually cannot be neglected
and interaction-induced delocalization
can compete with disorder-induced localization effects 
\cite{clement2005,fort2005,clement2006}. 
Generally,
the interplay between the kinetic
energy, the interactions and the disorder is still a open question that
has motivated many works 
\cite{giamarchi1988,fisher1989,dorokhov1990,shepelyanski1994,graham2005}. 
It is clear from Eq.~(\ref{eq:TF0}) that, in the TF regime ($\sigmas \gg \xi$), 
where the interaction forces the wave function to adapt
to the random potential, a BEC will not localize.
Indeed, if $V(\vect{r})$ is a {\it homogeneous} random function\footnote{In
this context, the term `homogeneous' means that all
local statistical properties of the random potential are independent
of the position.} \cite{lifshits1988},
so is the BEC wave function, $\psi$, which therefore, cannot 
decay at large distances.
This has been confirmed in recent experiments \cite{clement2005,fort2005,clement2006}. 
The question thus arises as to understand whether, as a naive transcription 
of the Ioffe-Regel criterion \cite{ioffe1960} would suggest, localization
can happen when $\sigmas < \xi$.

In this paper, we show that this criterion is actually not sufficient for BECs at equilibrium
if the interactions are non-negligible (\ie if $\xi \ll L$, where $L$ is the size of the system). 
We indeed show that interaction-induced delocalization still overcomes localization 
effects even when $\xi \gg \sigmas$. In fact, due to the {\it smoothing} of the random 
potential \cite{lee1990}, the effect of disorder turns out to be reduced when 
$\xi / \sigmas$ increases.

In the following, we develop a general formalism based on perturbation theory 
(see section~\ref{sec:smooth}) 
to determine the BEC wave function in any given weak potential, $V(\vect{r})$, 
for an arbitrary ratio $\sigmas/\xi$. 
We find that the BEC density, $|\psi|^2$, is still given by Eq.~(\ref{eq:TF0}), 
except that the potential $V(\vect{r})$ has to be replaced by a 
{\it smoothed potential}, $\widetilde{V}(\vect{r})$. 
We derive an exact formula for the smoothed potential up to first order in the 
perturbation series. 
We then apply our results to the case where $V(\vect{r})$ is a 1D homogeneous
random potential (see section~\ref{sec:random}) and derive the statistical properties
of the smoothed random potential, $\widetilde{V}(\vect{r})$. From this, we conclude
that an interacting BEC remains delocalized, even for $\xi \gg \sigmas$ (if $\xi \ll L$).

%%%%%%%%%%%%%%%%%%%%%%%%%%%%%%%%%%%%%%%%%%%%%%%%%%%%%%%%%%%%%%%%%%%%%%
\section{Smoothing effect in interacting Bose-Einstein condensates}
\label{sec:smooth}

Consider a low-temperature Bose gas in $d$ dimensions with contact atom-atom 
interactions, $\gdD{d} \delta^{(d)}(\vect{r})$, where $\gdD{d}$ is the $d$-dimensional 
interaction parameter.
In 3D geometries, $\gdD{3}=4\pi\hbar^2\asc/m$, where $\asc$ is the scattering length
\cite{pitaevskii2004}, and $m$ is the atomic mass. 
Low dimensional geometries (1D or 2D) can be realized in ultracold atomic 
samples using a tight radial confinement, so that the radial wave function is frozen
to zero-point oscillations in the form $\varphi^0_\perp (\vect{r}_\perp)$, where
$\vect{r}_\perp$ is the radial coordinate vector. In this case,
$\gdD{d}=\gdD{3} \int \textrm{d}\vect{r}_\perp\ |\varphi^0_\perp (\vect{r}_\perp)|^4$.
For instance, one finds $\gdD{1}=2\hbar\omega_\perp\asc$, 
for a 2D harmonic radial confinement of frequency $\omega_\perp$.
In addition, the Bose gas is assumed to be subjected to a given potential, $V(\vect{r})$,
with a typical amplitude $\Vs$ and a typical variation scale $\sigmas$. 
Possibly, the potential, $V(\vect{r})$, may have various length scales.
In this case, we assume that $\sigmas$ is the smallest.
Assuming weak interactions, \ie $\overline{n}^{2/d-1} \gg m \gdD{d} / \hbar^2$,
where $\overline{n}$ the mean density
\cite{petrov2000,petrov2004}, we treat the BEC in the mean-field 
approach \cite{pitaevskii2004} and we use the Gross-Pitaevskii equation (GPE):
%+++++++++++++++++++++++++++++++++++++++++%
\begin{equation}
\mu \psi (\vect{r}) = \left[ \frac{-\hbar^2 \vect{\nabla}^2 }{2m} + V (\vect{r}) + 
             \gdD{d} |\psi (\vect{r})|^2 \right] \psi (\vect{r}) ~,
\label{GPE}
\end{equation}
%+++++++++++++++++++++++++++++++++++++++++% 
where $\mu$ is the BEC chemical potential, and where the wave function, $\psi$, is normalized
to the total number of condensed atoms, $\int \textrm{d}\vect{r}\ |\psi (\vect{r})|^2 = N$.
Note that $\psi$ minimizes the $N$-body energy functional so that $\psi$ is
necessary a real function (up to a non-physical uniform phase).
In 1D and 2D geometries and in the absence of trapping, 
no true BEC can exist due to significant long-wavelength 
phase fluctuations \cite{popov1983}. In this case, no  macroscopic wave function,
$\psi$, can be defined. However, because density fluctuations are strongly
suppressed in the presence of interactions, the Bose gas forms a
{\it quasicondensate} \cite{popov1983} and the density, $n$, can be treated
as a classical field. It turns out that $\sqrt{n}$ is governed by Eq.~(\ref{GPE}).
Therefore, even though we only refer to BEC wave functions in the following, 
our formalism also applies to quasicondensates, after replacing $\psi$ by $\sqrt{n}$.

	%%%%%%%%%%%%%%%%%%%%%%%%%%%%%%%%%%%
	\subsection{The Thomas-Fermi regime}
\label{sec:smooth.TF}

In the simplest situation, the healing length of the BEC is much smaller than the typical length 
scale of the potential ($\xi \ll \sigmas$). 
Therefore, the kinetic energy term in the GPE~(\ref{GPE}) is small and the BEC 
density, $|\psi|^2$, simply follows the 
spatial modulations of the potential:
%+++++++++++++++++++++++++++++++++++++++++%
\begin{eqnarray}
& & |\psi (\vect{r})|^2 = [\mu - V(\vect{r})] / \gdD{d} 
\hspace{0.5cm} \textrm{for~} \mu > V(\vect{r}) \nonumber \\
\textrm{and} & & |\psi (\vect{r})|^2 = 0
\hspace{2.6cm} \textrm{otherwise}.
\label{TF}
\end{eqnarray}
%+++++++++++++++++++++++++++++++++++++++++%
This corresponds to the TF regime.
Note that for $\Vs \ll \mu$, one has
%+++++++++++++++++++++++++++++++++++++++++%
\begin{equation}
\psi (z) \simeq \psi_0 - \frac{V(\vect{r}) \psi_0}{2\mu} ~,
\label{dpsiTF}
\end{equation}
%+++++++++++++++++++++++++++++++++++++++++%
with $\psi_0 = \sqrt{\mu/\gdD{d}}$ being the BEC wave function at $V(\vect{r})=0$.
Therefore, the BEC wave function itself follows the modulations of the potential
$V(\vect{r})$.

	%%%%%%%%%%%%%%%%%%%%%%%%%%%%%%%%%%%
	\subsection{Beyond the Thomas-Fermi regime: the smoothing effect}
\label{sec:smooth.smooth}
The situation changes when the healing length is of the order of, or 
larger than, the typical length scale of the potential ($\xi > \sigmas$).
Indeed, the kinetic contribution limits the smallest variation length of the 
spatial modulations of a BEC wave function to a finite value of the order
of the healing length \cite{pitaevskii2004}. 
Therefore, the BEC can only follow modulations 
of the potential on a length scale typically larger than $\xi$ and 
Eq.~(\ref{TF}) no longer holds.

For a weak amplitude of the 
potential\footnote{A precise condition for the validity of the 
perturbative approach will be given later [see Eq.~(\ref{perturbvalid0})].}, 
we can use perturbation theory techniques. We thus write
the BEC wave function as $\psi (\vect{r}) = \psi_0 + \delta\psi (\vect{r})$ where
we assume that $\delta\psi \ll \psi_0$, and $\psi_0$ is the zeroth-order solution
of the GPE~(\ref{GPE}):
%+++++++++++++++++++++++++++++++++++++++++%
\begin{equation}
\mu \psi_0 = -\frac{\hbar^2}{2m}\vect{\nabla}^2 \psi_0 + \gdD{d} \psi_0^{3} ~.
\label{GPEzeroorder}
\end{equation}
%+++++++++++++++++++++++++++++++++++++++++%
Since the BEC is homogeneous at zeroth-order, one has $\psi_0 = \sqrt{\mu/\gdD{d}}$.
Then, the first order term of the perturbation series is given by 
%+++++++++++++++++++++++++++++++++++++++++%
\begin{equation}
-\frac{\hbar^2}{2m} \vect{\nabla}^2 (\delta \psi) 
- \left[ \mu - 3\gdD{d} \psi_0^2 \right] \delta \psi
= - V (\vect{r}) \psi_0 ~.
\label{GPEoneorder}
\end{equation}
%+++++++++++++++++++++++++++++++++++++++++%
Since $\mu -3\gdD{d} \psi_0^2 = -2\mu$, we are left with the equation
%+++++++++++++++++++++++++++++++++++++++++%
\begin{equation}
-\frac{\xi^2}{2} \vect{\nabla}^2 (\delta \psi) 
+ \delta \psi
= - \frac{V (\vect{r}) \psi_0}{2\mu} ~,
\label{GPEoneorder2}
\end{equation}
%+++++++++++++++++++++++++++++++++++++++++%
where $\xi = \hbar / \sqrt{2m\mu}$ is the healing length of the BEC. 
We straightforwardly find the solution of Eq.~(\ref{GPEoneorder2}),
which reads
%+++++++++++++++++++++++++++++++++++++++++%
\begin{equation}
\delta\psi (\vect{r}) = 
- \int\textrm{d}\vect{r'}\ \gecr (\vect{r}-\vect{r'}) \frac{V (\vect{r'}) \psi_0}{2\mu} ~,
\label{dpsiLDA}
\end{equation}
%+++++++++++++++++++++++++++++++++++++++++%
where $\gecr (\vect{r})$ is the Green function of Eq.~(\ref{GPEoneorder2}),
defined as the solution of 
%+++++++++++++++++++++++++++++++++++++++++%
\begin{equation}
\left[-\frac{\xi^2}{2} \vect{\nabla}^2 + 1\right] \gecr (\vect{r}) 
= \delta^{(d)}(\vect{r}) ~,
\label{green}
\end{equation}
%+++++++++++++++++++++++++++++++++++++++++%
or equivalently, in Fourier space
%+++++++++++++++++++++++++++++++++++++++++%
\begin{equation}
\left[\frac{\xi^2}{2} |\vect{k}|^2 + 1 \right] \widehat{\gecr}(\vect{k}) 
= 1/(2\pi)^{d/2} ~,
\label{greenFT}
\end{equation}
%+++++++++++++++++++++++++++++++++++++++++%
where, $\widehat{\gecr}(\vect{k})=\frac{1}{(2\pi)^{d/2}}\ \int \textrm{d}\vect{r}\ 
\gecr(\vect{r}) \textrm{e}^{-\textrm{i}\vect{k} \cdot \vect{r}}$ is the Fourier transform
of $\gecr$. In contrast to the case of single particles, the Green function,
$\widehat{\gecr}(\vect{k})$, has no singularity point so that the perturbative approach
can be safely applied for any wave vector $\vect{k}$.

The explicit formula for the Green function, $\gecr$, depends on the dimension of the system.
After some simple algebra, we find
%+++++++++++++++++++++++++++++++++++++++++%
\begin{eqnarray}
\textrm{in~} 1\textrm{D}, &\hspace{0.2cm}& \gecr (z) = \frac{1}{\sqrt{2}\xi} \exp\left(-\frac{|z|}{\xi / \sqrt{2}}\right)
\label{ge1D} \\
\textrm{in~} 2\textrm{D}, && \gecr (\vect{\rho}) = \frac{1}{\pi\xi^2} \textrm{K}_0 \left(+\frac{|\rho|}{\xi / \sqrt{2}}\right)
\label{ge2D} \\
\textrm{in~} 3\textrm{D}, && \gecr (\vect{r}) = \frac{1}{2\pi \xi^2 |\vect{r}|} \exp\left(-\frac{|\vect{r}|}{\xi / \sqrt{2}}\right) ~.
\label{ge3D}
\end{eqnarray}
%+++++++++++++++++++++++++++++++++++++++++%
where $\textrm{K}_0$ is the modified Bessel function.
Finally, up to first order in the perturbation series, the BEC wave function reads
%+++++++++++++++++++++++++++++++++++++++++%
\begin{equation}
\psi (\vect{r}) \simeq \psi_0 - \frac{\widetilde{V} (\vect{r}) \psi_0}{2\mu}
\label{dpsiLDAsol}
\end{equation}
%+++++++++++++++++++++++++++++++++++++++++%
with
%+++++++++++++++++++++++++++++++++++++++++%
\begin{equation}
\widetilde{V} (\vect{r}) = 
\int \textrm{d}\vect{r'} \ \gecr (\vect{r'}) \Vopt(\vect{r}-\vect{r'}) ~.
\label{vopteff}
\end{equation}
%+++++++++++++++++++++++++++++++++++++++++%

Interestingly enough, the Green function in any dimension shows a exponential decay,
with a typical attenuation length, $\xi$, and is normalized to 
unity\footnote{This property
follows directly from the definition~(\ref{greenFT}) of the Green function.
Indeed, 
$\int \textrm{d}\vect{r}\ \gecr (\vect{r}) = (2\pi)^{d/2} \widehat{\gecr}(\vect{k}=0) = 1$.}, 
$\int \textrm{d}\vect{r}\ \gecr (\vect{r}) = 1$.
Therefore, $\gecr (\vect{r})$ can be seen as a {\it smoothing function} 
with a typical width $\xi$.
Indeed, it should be noted that Eq.~(\ref{dpsiLDAsol}) is similar to Eq.~(\ref{dpsiTF}),
except that the potential $V(\vect{r})$ which is relevant in the case
$\xi \ll \sigmas$, changes to the potential $\widetilde{V}(\vect{r})$
for $\xi > \sigmas$. 
The potential $\widetilde{V}(\vect{r})$ is a convolution of $V(\vect{r})$ with a
function which has a typical width $\xi$ and thus corresponds to
a {\it smoothed potential} with an amplitude smaller than $\Vs$.
In addition, if $\sigmas$ corresponds to the width of the correlation function of
a random potential, $V$, the correlation length of the smoothed random potential, 
$\widetilde{V}$, is of the order of $\max (\sigmas$,$\xi)$ 
[for details, see section~\ref{sec:random}]\footnote{In contrast, for example in the case of a 
deterministic periodic potential, 
$V(z)=\Vs \cos\left(kz\right)$, the variation scale, $\sigmas=2\pi/k$, corresponds
to the period of the potential, and we find 
$\widetilde{V}(z)=\frac{\Vs\cos\left(kz\right)}{1+k^2\xi^2}$.
The potential is indeed smoothed as the amplitude of $\widetilde{V}$
is smaller than that of $V$.
Nevertheless, the period of the smoothed potential, $\widetilde{V}$, is the same as that of 
the bare potential, $V$.}.

Note that, for $\xi \ll \sigmas$, $\gecr(\vect{r})$ can be approximated by
$\delta^{(d)}(\vect{r})$ in Eq.~(\ref{vopteff}),
and $\widetilde{V}(\vect{r}) \simeq V(\vect{r})$. 
We thus recover the results of section~\ref{sec:smooth.TF},
valid for the TF regime.

The validity condition of the perturbation approach directly follows from
Eq.~(\ref{dpsiLDAsol}):
%+++++++++++++++++++++++++++++++++++++++++%
\begin{equation}
\widetilde{V} (\vect{r}) \ll \mu ~.
\label{perturbvalid0}
\end{equation}
%+++++++++++++++++++++++++++++++++++++++++%
Note that if $\xi \gg \sigmas$, the potential can be significantly smoothed
so that the above condition can be less restrictive than the {\it a priori
condition}, $V(\vect{r}) \ll \mu$.

The results of this section show that the potential, $V(\vect{r})$, can be
significantly smoothed in interacting BECs. We stress that this applies to
any kind of potentials provided that $\xi \ll L$ and $\widetilde{V}(\vect{r}) \ll \mu$.
In the next section, we present an illustration of the smoothing effect
in the case of a random potential.

%%%%%%%%%%%%%%%%%%%%%%%%%%%%%%%%%%%%%%%%%%%%%%%%%%%%%%%%%%%%%%%%%%%%%%
\section{Application to a trapped interacting Bose-Einstein condensate 
in a 1D random potential}
\label{sec:random}

	%%%%%%%%%%%%%%%%%%%%%%%%%%%%%%%%%%%
	\subsection{Trapped 1D Bose-Einstein condensate in a random potential}
\label{sec:random.bec}

In this section, we consider a 1D Bose gas subjected to a weak
homogeneous random potential, $V(z)$, with a vanishing average value 
($\langle V \rangle = 0$),
a standard deviation, $\Vs$, and a spatial correlation length, $\sigmas$,
significantly smaller than the size of the system.
In addition, we assume that the gas is trapped in a confining harmonic 
trap\footnote{All results also apply if there is no trapping. 
In this case, all zeroth-order terms simply do not depend on $z$.}, 
$V_\textrm{\tiny h}(z)=m\omega^2z^2/2$ as in almost all current experiments 
on disordered BECs \cite{lye2005,clement2005,fort2005,clement2006}.
We consider a situation such that $\hbar\omega \ll n\gdD{1} \ll \hbar^2n^2/m$,
\ie the Bose gas lies in the mean-field regime, and in the absence of disorder,
the interactions dominate over the kinetic energy\footnote{This corresponds to the 
usual TF regime for confined BECs in the absence of disorder
\cite{pitaevskii2004}. However, 
no restriction is imposed for the ratio $\sigmas/\xi$, so that the BEC can be out of the 
TF regime as defined in section~\ref{sec:intro}.}. 
The situation mimics the experimental conditions of Ref.~\cite{clement2005,clement2006}.
The presence of the harmonic confinement introduces a low-momentum cut-off for the
phase fluctuations so that the 1D Bose gas forms a true condensate at low temperatures
\cite{petrov2000,petrov2004}.
In this case, the BEC wave function is
%+++++++++++++++++++++++++++++++++++++++++%
\begin{equation}
\psi_0 = \sqrt{\mu_0 (z)/\gdD{1}} ~,
\label{psiTF}
\end{equation}
%+++++++++++++++++++++++++++++++++++++++++%
where $\mu_0 (z)=\mu-m\omega^2z^2/2$ is the local chemical potential.
This corresponds to an inverted parabolic density profile with a half-size,
$\LTF=\sqrt{2\mu / m\omega^2}$,
where the chemical potential is
$\mu=\muTF = \frac{\hbar \omega}{2}\left(\frac{3Nm\goned l}{2\hbar^2}\right)^{2/3}$,
with $l=\sqrt{\hbar/m\omega}$ being the extension of the ground state of the
harmonic oscillator. 

As $\LTF \gg (\xi,\sigmas)$, it is legitimate to use the local
density approximation (LDA) \cite{pitaevskii2004}, \ie in a region
significantly smaller than $\LTF$, the quantities
$\psi_0$ and $\mu_0$ can be considered as uniform. 
We can thus directly apply the results of section~\ref{sec:smooth.smooth}.
From Eqs.~(\ref{dpsiLDAsol})-(\ref{psiTF}), we immediately find that
%+++++++++++++++++++++++++++++++++++++++++%
\begin{equation}
n (z) \simeq n_0 (z) - \frac{\widetilde{V} (z)}{\gdD{1}} ~,
\label{densLDAsol}
\end{equation}
%+++++++++++++++++++++++++++++++++++++++++%
where
%+++++++++++++++++++++++++++++++++++++++++%
\begin{equation}
\widetilde{V} (z) = 
\int \textrm{d}z' \ 
\frac{\exp\left(\frac{-|z'|}{\xi_0 (z)/\sqrt{2}}\right)}{\sqrt{2}\xi_0 (z)} V(z-z') ~,
\label{vopteffLDA}
\end{equation}
%+++++++++++++++++++++++++++++++++++++++++%
is the smoothed 
potential, with $\xi_0 (z)=\hbar/\sqrt{2m\mu_0(z)}$
being the local healing length. The density profile of the BEC is thus expected to
follow the modulations of a {\it smoothed random potential}.

Note that the total number of condensed atoms is 
$N=\int \textrm{d}z\ |\sqrt{n_0 (z)}+\delta\psi|^2 
\simeq \int \textrm{d}z\ \left( n_0 (z) - \tVopt (z)/\goned \right)$ 
up to first
order in $\tVopt/\mu$. Since $\langle \tVopt \rangle = 0$, one has 
$\langle N \rangle \simeq \int \textrm{d}z\ n_0 (z)$,
owing to the assumed self-averaging property of the potential
\cite{lifshits1988}. In addition, we have $\mu = \muTF$.

%-----------------------------------------%
\begin{figure}[t!]
\begin{center}
\infig{25em}{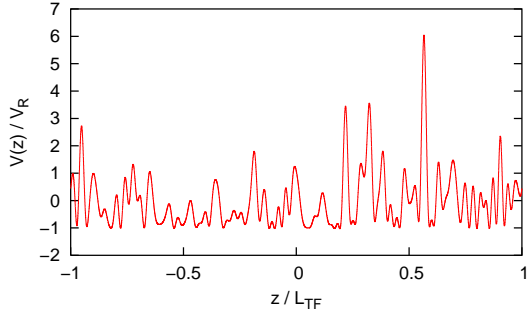}
\end{center}
\caption{(color online)
Example of the realization of a speckle random potential with 
$\sigmas \simeq 10^{-2} \LTF$.
} 
\label{potential}
\end{figure}
%-----------------------------------------%

%-----------------------------------------%
\begin{figure}[t!]
\begin{center}
\infig{27.em}{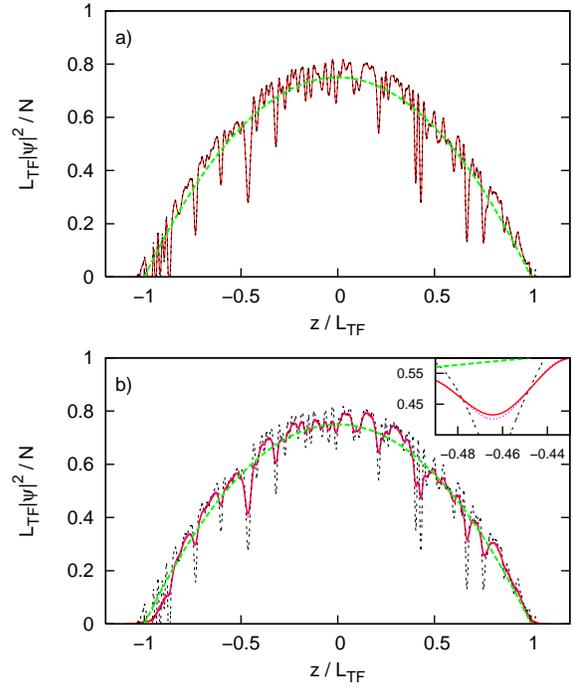}
\end{center}
\caption{(color online)
Density profiles of a BEC confined in a combined harmonic plus random potential
($\Vs=0.1\mu$, $\sigmas=7.5\times 10^{-3}\LTF$). 
The solid (red online) line corresponds to the numerically computed BEC
wave function; the dashed (green online) line is the TF profile in the absence
of disorder; and the black dotted line is a plot of the disordered TF
profile [Eq.~(\ref{TF})].
a) Case where the healing length at the trap center, $\xi$, is smaller than 
the correlation length
of the random potential: $\sigmas / \xi \simeq 10$. In this case, the
density profile follows the modulations of the random potential according to
Eq.~(\ref{TF}).
b) Opposite situation: $\sigmas / \xi \simeq 0.5$. In this case, the BEC density profile,
obtained numerically, significantly differs from Eq.~(\ref{TF}), but can hardly be distinguished
from Eq.~(\ref{densLDAsol}) [also plotted in Fig.~\ref{densityhealing}b) as a dotted (purple online)
line]. The inset shows a magnification of the plot in a very small region of the BEC.
} 
\label{densityhealing}
\end{figure}
%-----------------------------------------% 

We now compare our predictions to the exact solutions of the GPE~(\ref{GPE}) 
as obtained numerically.
For the sake of concreteness, we consider a speckle random potential \cite{goodman} 
similar
to the one used in the recent experiments \cite{lye2005,clement2005,fort2005,clement2006}
(see Fig.~\ref{potential}).
Briefly, a speckle pattern consists in a random intensity distribution and is
characterized by its statistical properties. 
First, the single-point amplitude distribution is a negative exponential 
%+++++++++++++++++++++++++++++++++++++++++%
\begin{eqnarray}
& & \mathcal{P}[V(z)] = \frac{\exp [ -(V(z)+\Vs)/\Vs ]}{|\Vs|} 
\textrm{~~~~if~} \frac{V(z)}{\Vs} \ge -1 \nonumber \\
\textrm{and} & & \mathcal{P}[V(z)] = 0 \label{probaintens}
\textrm{~~~~otherwise} ~,
\end{eqnarray}
%+++++++++++++++++++++++++++++++++++++++++%
corresponding to the average value 
$\langle \Vopt \rangle = 0$ 
and the standard deviation
$\Delta \Vopt=\sqrt{\langle [\Vopt(z)-\langle \Vopt \rangle]^2 \rangle} = |\Vs|$. 
Second, the spatial correlations are characterized by the autocorrelation function
$\corr(\Delta z) = \langle \Vopt(\Delta z)\Vopt(0) \rangle$ 
which correlation length is denoted $\sigmas$ and
can be chosen at will \cite{clement2006,goodman}.
For the numerical calculations, we numerically generate 
a 1D speckle pattern using a method similar to the one described in 
Ref.~\cite{huntley1989}
in 1D and corresponding to the correlation function  
%+++++++++++++++++++++++++++++++++++++++++%
\begin{equation}
\corr(\Delta z) = \Vs^2 \left|\sinc(\sqrt{3}\Delta z/\sqrt{2}\sigmas)\right|^2 ~,
\end{equation}
%+++++++++++++++++++++++++++++++++++++++++%
where $\sinc (x)=\sin (x)/x$.
For the sake of simplicity, it is useful to approximate $\corr (z)$ to
a Gaussian function (see for example section~\ref{sec:random.stat}). Up to
second order in $\Delta z/\sigmas$, we have
$\corr (\Delta z) \simeq \Vs^2 \exp(-\Delta z^2/2\sigmaG^2)$.

Numerical solutions of the GPE~(\ref{GPE}) are presented in Fig.~\ref{densityhealing} 
for two values of the ratio $\sigmas/\xi$, where $\xi$ is the BEC healing length
at the trap center.
In the first case (Fig.~\ref{densityhealing}a), we have $\xi \ll \sigmas$, and the
density simply 
follows the modulations of the bare random potential,
according to Eq.~(\ref{TF}).
In the second case (Fig.~\ref{densityhealing}b), we have $\xi > \sigmas$,
and as expected, the BEC wave function does not follow the modulations of the
bare random potential $V(z)$ but actually follows smoother modulations of
the smoothed potential $\widetilde{V}(z)$. Figure~\ref{densityhealing}b (and the inset)
shows that the numerically computed density can hardly be distinguished from 
Eq.~(\ref{densLDAsol}).
This supports the validity of our approach.

	%%%%%%%%%%%%%%%%%%%%%%%%%%%%%%%%%%%
	\subsection{Statistical properties of the smoothed random potential}
\label{sec:random.stat}

\begin{figure}[t!]
\begin{center}
\infig{29.em}{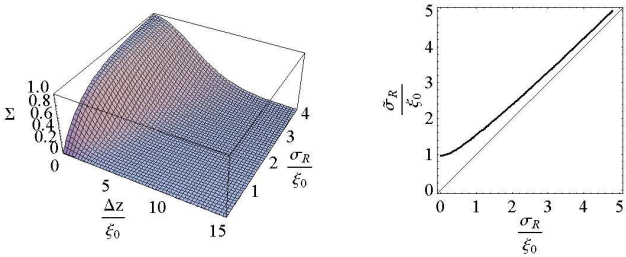}
\end{center}
\caption{(color online)
Left: Plot of the correlation function $\Sigma \left( \frac{\sigmaG}{\xi_0},\frac{\Delta z}{\xi_0}\right)$.
Right: Width at $1/\sqrt{e}$ of the normalized correlation function 
$\Sigma \left( \frac{\sigmaG}{\xi_0},\frac{\Delta z}{\xi_0}\right)/
\Sigma \left( \frac{\sigmaG}{\xi_0},0\right)$.
} 
\label{sigma}
\end{figure}

It is useful to compute the statistical properties of the smoothed
random potential $\widetilde{V} (z)$ as they will be imprinted on the BEC density
profile according to Eq.~(\ref{densLDAsol}). 
From Eq.~(\ref{vopteff}), we immediately find that,
(i) $\widetilde{V} (z)$ is a random {\it homogeneous} potential,
(ii) the average value of $\widetilde{V}$ vanishes,
%+++++++++++++++++++++++++++++++++++++++++%
\begin{equation}
\langle \widetilde{V} \rangle = \langle V \rangle = 0  ~,
\label{meantilde}
\end{equation}
%+++++++++++++++++++++++++++++++++++++++++%
and (iii) the correlation function of $\widetilde{V}$ is given by
%+++++++++++++++++++++++++++++++++++++++++%
\begin{equation}
\tcorr_z (\Delta z) = \int \textrm{d}u \textrm{d}v \ \corr[\Delta z+(v-u)] 
                \gecr_{z} (u) \gecr_{z+\Delta z}  (v) ~,
\label{corrtilde}
\end{equation}
%+++++++++++++++++++++++++++++++++++++++++%
where $\corr(\Delta z)=\langle \Vopt(\Delta z) \Vopt(0) \rangle$ 
is the correlation function of the bare potential $\Vopt$
and $G_z (u)$ is given by Eq.~(\ref{ge1D}) with $\xi$ replaced by $\xi_0 (z)$.
In the following, we assume that $\Delta z \ll \LTF$ so that 
$G_z \simeq G_{z+\Delta z}$ and we omit the subscripts.
Assuming for simplicity a Gaussian correlation function for the
bare random potential,
$\corr(\Delta z) ~\simeq~ \Vs^2 \exp(-\Delta z^2/2\sigmaG^2)$, we find after some algebra 
%+++++++++++++++++++++++++++++++++++++++++%
\begin{equation}
\tcorr (\Delta z) = \Vs^2 \ \Sigma \left( \frac{\sigmaG}{\xi_0},\frac{\Delta z}{\xi_0}\right) ~,
\label{corrtildegauss1}
\end{equation}
%+++++++++++++++++++++++++++++++++++++++++%
with
%+++++++++++++++++++++++++++++++++++++++++%
\begin{widetext}
\begin{eqnarray}
\Sigma(\sigmaGline,\overline{\Delta z}) & = & \sigmaGline^2 \exp\left(-\frac{\overline{\Delta z}^2}{2\sigmaGline^2}\right) \nonumber \\
&& + \frac{\sqrt{\pi}}{4} \sigmaGline \left( 1 - 2 \sigmaGline^2 - \sqrt{2} \ \overline{\Delta z} \right)  
                       \exp\left(\sigmaGline^2 + \sqrt{2} \ \overline{\Delta z}\right)
% \\
% &&		       \phantom{+} \times 
                       \erfc\left( \frac{2\sigmaGline^2 + \sqrt{2} \ \overline{\Delta z}}{2\sigmaGline} \right) \label{corrtildegauss2} \\
&& + \frac{\sqrt{\pi}}{4} \sigmaGline \left( 1 - 2 \sigmaGline^2 + \sqrt{2} \ \overline{\Delta z} \right)  
                       \exp\left(\sigmaGline^2 - \sqrt{2} \ \overline{\Delta z}\right) \nonumber 
% \\
% &&		       \phantom{+} \times 
                       \erfc\left( \frac{2\sigmaGline^2 - \sqrt{2} \ \overline{\Delta z}}{2\sigmaGline} \right) \nonumber
\end{eqnarray}
\end{widetext}
%+++++++++++++++++++++++++++++++++++++++++%
where $\sigmaGline=\sigmaG/\xi_0$, $\overline{\Delta z}=\Delta z/\xi_0$ and 
$\erfc~(x)~=~\frac{2}{\sqrt{\pi}}\int_{x}^{\infty}\textrm{d}t~e^{-t^2}$
is the complementary error function. 
The correlation function 
$\Sigma \left( \frac{\sigmaG}{\xi_0},\frac{\Delta z}{\xi_0}\right)$
is plotted in Fig.~\ref{sigma}. 

%-----------------------------------------%
%-----------------------------------------% 

This function $\Sigma$ clearly decreases with $\sigmaG/\xi_0$, indicating the onset
of an increasing smoothing effect.
At $\Delta z=0$, we have a simple asymptotic expression
for $\sigmaG \gg \xi_0$:
%+++++++++++++++++++++++++++++++++++++++++%
\begin{equation}
\Sigma(\sigmaG/\xi_0,0) \simeq 1 - \left(\frac{\xi_0}{\sigmaG}\right)^2,~~
\sigmas \gg \xi_0 ~.
\label{eq:sigmaGasym1}
\end{equation}
%+++++++++++++++++++++++++++++++++++++++++%
So, as expected, $\Sigma(\sigmaG/\xi_0,0) \rightarrow 1$ 
as $\sigmaG/\xi_0 \rightarrow \infty$, \ie the random potential is hardly smoothed.
For $\sigmaG \ll \xi_0$:
%+++++++++++++++++++++++++++++++++++++++++%
\begin{equation}
\Sigma(\sigmaG/\xi_0,0) \simeq \frac{\sqrt{\pi}}{2} 
\frac{\sigmaG}{\xi_0},~~
\sigmas \ll \xi_0 ~.
\label{eq:sigmaGasym2}
\end{equation}
%+++++++++++++++++++++++++++++++++++++++++%
So,
$\Sigma(\sigmaG/\xi_0,0) \rightarrow 0$ 
as $\sigmaG/\xi_0 \rightarrow 0$, \ie the amplitude of the smoothed
random potential is significantly reduced compared to the amplitude of 
the bare random potential.
Generally speaking, from Eq.~(\ref{corrtildegauss1}), we have
$\langle \tVopt^2 \rangle = \tcorr (0) = \Vs^2 \Sigma (\sigmaG/\xi_0,0)$.
It follows that $\langle \tVopt^2 \rangle$ is an increasing function of
$\sigmaG/\xi_0$ and  that $\langle \tVopt^2 \rangle\leq \Vs^2$.
This is consistent
with the idea of a {\it smoothing} of the random potential.

In addition, the correlation length, $\tsigmas$, of the smoothed random
potential, $\widetilde{V}$, is given by the width at $1/\sqrt{e}$ of the function
$\Delta z \rightarrow \Sigma (\sigmaGline/\xi_0,\Delta z/\xi_0)$. 
At $\sigmas \gg \xi_0$, the smoothing is weak and $\tsigmas \simeq \sigmas$.
For $\sigmas < \xi_0$, the smoothing is significant, so that $\tsigmas$
saturates at $\tsigmas \simeq \xi_0$, as expected. Roughly speaking, 
we have $\tsigmas \sim \max (\sigmas,\xi_0)$ [see Fig.~\ref{sigma}].

	%%%%%%%%%%%%%%%%%%%%%%%%%%%%%%%%%%%
	\subsection{Effect of disorder in interacting Bose-Einstein condensates}
\label{sec:random.prop}

We finally discuss the properties of the BEC wave function in the presence of
disorder. It follows from Eq.~(\ref{densLDAsol}) that
the BEC density follows the modulations of a random potential $\widetilde{V}$.
In the TF regime ($\xi \ll \sigmas$), $\widetilde{V} \simeq V$, while
when $\xi > \sigmas$, $\widetilde{V}$ is smoothed.
Since $\widetilde{V}$ is a homogeneous random potential, there is no decay of the 
wave function. In particular, Anderson localization does not occur,
even for $\xi \gg \sigmas$. 
In the case when $\xi>\sigmas$, it turns out that the BEC density is actually less 
affected by the random potential than in the TF regime ($\xi \ll \sigmas$).
This is in striking contrast with the case of non-interacting
particles where localization effects are usually stronger at low energy
\cite{lifshits1988}.

More quantitatively, using the statistical properties of the smoothed random potential, $\widetilde{V}$,
one can easily compute the fluctuations 
$\Delta n (z) = \sqrt{\langle [n(z)-n_0(z)]^2 \rangle}$ 
of the BEC density around the average value 
$n_0(z)=[\mu-m\omega^2 z^2/2]/\goned$. From Eq.~(\ref{densLDAsol}),
we find $\Delta n^2 \simeq \tcorr (0)/\goned^2$. Note that
$\Delta n^2$ depends on the displacement from the trap center
through the dependence of $\xi_0$ on $z$.
At the trap center, we find
%+++++++++++++++++++++++++++++++++++++++++%
\begin{equation}
\Delta n_\textrm{c} = \frac{\Vs}{\goned}
\sqrt{\Sigma (\sigmaG/\xi,0)} ~.
\label{modhealing}
\end{equation}
%+++++++++++++++++++++++++++++++++++++++++%
We recall that $\xi = \xi_0 (0) = \hbar/\sqrt{2m\mu}$ is the BEC healing length
in the trap center.

\begin{figure}[t!]
\begin{center}
\infig{27em}{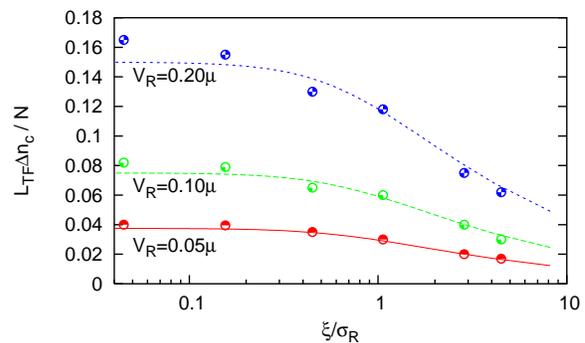}
\end{center}
\caption{(color online)
Amplitude of the fluctuations of the BEC density at the trap center, $\Delta n_\textrm{c}$, 
in the combined harmonic 
plus random potential as a function of the ratio of the healing length to the 
correlation length of the disorder for several amplitudes of the random
potential. The dots correspond to exact numerical results in the Gross-Pitaevskii approach 
[Eq.~(\ref{GPE})]
and the lines show the theoretical prediction [Eq.~(\ref{modhealing})].
} 
\label{fluctdensity}
\end{figure}

We have numerically extracted the fluctuations of the density in the trap center,
according to the formula
$\Delta n_\textrm{c} \simeq \sqrt{\frac{1}{\LTF/2}\int_{-\LTF/4}^{+\LTF/4} \textrm{d}z\
\left[n(z)-n_0(z)\right]^2}$. This provides a good estimate of $\Delta n_\textrm{c}$
as $\xi_0 (z)$ changes by less than $3\%$ in the range $[-\LTF/4,+\LTF/4]$.
As shown in Fig.~\ref{fluctdensity}, the numerical results
perfectly agree with 
Eq.~(\ref{modhealing}) over a large range of the ratio $\xi/\sigmas$.
The numerical calculations are performed for the speckle potential 
described in section~\ref{sec:random.stat} and no fitting 
parameter has been used.
In addition, note that we have used a single realization of the random
potential for each point in Fig.~\ref{fluctdensity}.
Averaging over disorder turned out to have little importance, since the random
potential is almost self-averaging in the range $[-\LTF/4,+\LTF/4]$, if $\sigmas \ll \LTF$.

%-----------------------------------------%
%-----------------------------------------%

Finally, we find from Eq.~(\ref{perturbvalid0}) that the perturbative approach 
that we have performed is valid 
whenever $\Delta n \ll n_0$, {\it \ie} whenever
%+++++++++++++++++++++++++++++++++++++++++%
\begin{equation}
\Vs \sqrt{\Sigma (\sigmaG/\xi_0,0)} \ll \mu ~.
\label{perturbvalid}
\end{equation}
%+++++++++++++++++++++++++++++++++++++++++%
Note that this effect is more restrictive in the trap center where
$\xi_0$ is minimum.

%%%%%%%%%%%%%%%%%%%%%%%%%%%%%%%%%%%%%%%%%%%%%%%%%%%%%%%%%%%%%%%%%%%%%%
\section{Conclusion}
\label{sec:concl}

In summary, we have presented an analytical technique, based on the perturbation
theory, to compute the static wave function of an interacting BEC subjected
to a weak potential. This applies to the case where both the healing length 
of the BEC ($\xi$) and the spatial scale of the potential ($\sigmas$) are 
much smaller than the size of the system ($L$), but whithout restriction for 
the ratio $\xi/\sigmas$.
In particular, we have shown that when the healing length
is larger than the space scale of the potential, the BEC is sensitive to a 
{\it smoothed potential} which can be determined within our framework.

Applying these results to the case of a 1D random potential, we have shown 
that the wave function of a static interacting BEC is delocalized, similarly 
as in the TF regime \cite{clement2005}.
This is confirmed by numerical calculations.
The results of this analysis show that, for an
interacting BEC at equilibrium, the larger the healing length, 
the smaller the perturbation induced by the disorder. 
It is worth noting that the conclusions of the present work hold for {\it static} BECs
in the {\it mean-field regime} and when the interaction energy dominates over
the kinetic energy in the absence of disorder, \ie when the healing
length is significantly smaller than the BEC half size ($\xi \ll L$). 
Going beyond the mean-field regime, it is interesting to study the
interplay of interactions, disorder and kinetic energy in a Bose gas
for interactions ranging from zero (where localization is expected) to the TF regime
(where the BEC is delocalized as shown in this work). This question is
addressed in Ref.~\cite{lugan2006}.

Finally, we note that the transport properties of a BEC can show significantly
different physics. 
For instance, localization has been studied in matter-wave beams \cite{paul2005} 
and in the expansion of an interacting BEC \cite{clement2005,fort2005,clement2006}.
In the latter two cases, localization indeed does occur although non-negligible interactions 
can modify the usual picture of localization \cite{paul2005,clement2005,clement2006}.

%%%%%%%%%%%%%%%%%%%%%%%%%%%%%%%%%%%%%%%%%%%%%%%%%%%%%%%%%%%%%%%%%%%%%%
\acknowledgements
We are indebted to G.~Shlyapnikov, A.~Aspect, M.~Lewenstein, D.~Gangardt, 
and P.~Bouyer for many stimulating discussions.
We thank D.~Cl\'ement and P.~Lugan for discussions and useful comments 
on the manuscript.
This work was supported by the Centre National de la Recherche Scientifique (CNRS), 
the D\'{e}l\'{e}\-gation G\'{e}n\'{e}rale de l'Armement,
the Agence Nationale de la Recherche (contract NTOR-4-42586),
the European Union (grants IST-2001-38863 and MRTN-CT-2003-505032),
and INTAS (Contract 211-855).
The Atom Optics group at LCFIO is a member of the Institut Francilien 
de Recherche sur les Atomes Froids (IFRAF).

%%%%%%%%%%%%%%%%%%%%%%%%%%%%%%%%%%%%%%%%%%%%%%%%%%%%%%%%%%%%%%%%%%%%%%

%%%%%%%%%%%%%%%%%%%%%%%%%%%%%%%%%%%%%%%%%%%%%%%%%%%%%%%%%%%%%%%%%%%%%%

\end{document}